\documentclass[aps,prl,reprint,superscriptaddress]{revtex4-2}

\usepackage{bm}
\usepackage{graphicx}

\begin{document}

\preprint{v3}

\title{Field Dispersion and Strong Coupling of Nuclear--Electron Spin Excitation in MnCO$_3$}


\author{Takahiko Makiuchi}
\email{takahiko.makiuchi@riken.jp}
\affiliation{RIKEN Center for Emergent Matter Science (CEMS), Wako 351–0198, Japan.}
\affiliation{Department of Applied Physics, the University of Tokyo, Tokyo 113-8656, Japan.}
\author{Takashi Kikkawa}
\affiliation{Department of Applied Physics, the University of Tokyo, Tokyo 113-8656, Japan.}
\author{Thanaporn Sichanugrist}
\affiliation{Department of Applied Physics, the University of Tokyo, Tokyo 113-8656, Japan.}
\affiliation{Department of Physics, the University of Tokyo, Tokyo 113-8656, Japan.}
\author{Junki Numata}
\affiliation{Department of Applied Physics, the University of Tokyo, Tokyo 113-8656, Japan.}
\author{Masaki Imai}
\affiliation{Advanced Science Research Center, Japan Atomic Energy Agency, Tokai 319-1195, Japan.}
\author{Hiroyuki Chudo}
\affiliation{Advanced Science Research Center, Japan Atomic Energy Agency, Tokai 319-1195, Japan.}
\author{Saburo Takahashi}
\affiliation{Advanced Institute for Materials Research, Tohoku University, Sendai 980-8577, Japan.}
\author{Eiji Saitoh}
\affiliation{RIKEN Center for Emergent Matter Science (CEMS), Wako 351–0198, Japan.}
\affiliation{Department of Applied Physics, the University of Tokyo, Tokyo 113-8656, Japan.}
\affiliation{Advanced Science Research Center, Japan Atomic Energy Agency, Tokai 319-1195, Japan.}
\affiliation{Advanced Institute for Materials Research, Tohoku University, Sendai 980-8577, Japan.}
\affiliation{Institute for AI and Beyond, the University of Tokyo, Tokyo 113-8656, Japan}


\date{\today}

\begin{abstract}
 Hybridized nuclear and electron spin excitation in a MnCO$_3$ crystal, a weakly-anisotropic antiferromagnet, has been investigated.
 In this material, the hyperfine interaction is strong enough to form a nuclear spin wave.
 We measure the microwave absorption by a bulk MnCO$_3$ and observe the dispersion representing strong frequency repulsion between electron and nuclear modes due to their hybridization, the signature of nuclear spin wave.
 Additionally, we observe that the nuclear spin resonance enters a nonlinear regime above a certain excitation power, attributed to the excitation of finite wavenumber nuclear spin waves.
\end{abstract}


\maketitle

In condensed matter, nuclear spins exist in isolation and interact with neighboring electron spins through the hyperfine interaction.
The hyperfine interaction has played a pivotal role in exploring electron spin states through nuclear magnetic resonance (NMR) \cite{abragam1961principles}.
Recently, the field of spintronics has been extended into the realm of the nucleus, a concept termed nuclear spintronics \cite{Shiomi2019, Kikkawa2021, rezende2022introduction, maekawa2023spin, kikkawa2023spin, chigusa2023dark}.
At the core of nuclear spintronics is the nuclear spin wave enabled by the Suhl--Nakamura interaction---an indirect interaction among nuclear spins facilitated by the exchange interaction between electron spins and the hyperfine interaction \cite{Suhl1958, Nakamura1958, deGennes1963}.
The nuclear spin wave operates within a megahertz frequency range \cite{Shiomi2019}, and its fluctuation persists even at very low temperatures \cite{Kikkawa2021}, which distinguishes it from the electron counterpart.
The conversion of nuclear spin waves into electron spin and charge currents may offer unique advantages such as extension of the frequency range and high coherence regime.

Materials suitable for nuclear spintronics include weakly-anisotropic antiferromagnets with strong hyperfine interactions \cite{Andrienko1991}, such as manganese carbonate MnCO$_3$ \cite{Shiomi2019, Kikkawa2021}.
The angular momenta in such materials can travel as electron and nuclear spin waves.
In this study, we clarify the field dispersion for the hybridized nuclear and electron spin excitations in MnCO$_3$ as a function of temperature by using broadband microwave spectroscopy.
The observed dispersions of electron and nuclear spins show the impact of the hybridization and a high cooperativity.
Examining the temperature dependence in these dispersions, we determine various material parameters of our MnCO$_3$ sample.
We also found a nonlinear regime in the nuclear spin wave dispersion indicating a formation of finite wavenumber nuclear spin waves.


\begin{figure}
 \includegraphics[width=80mm]{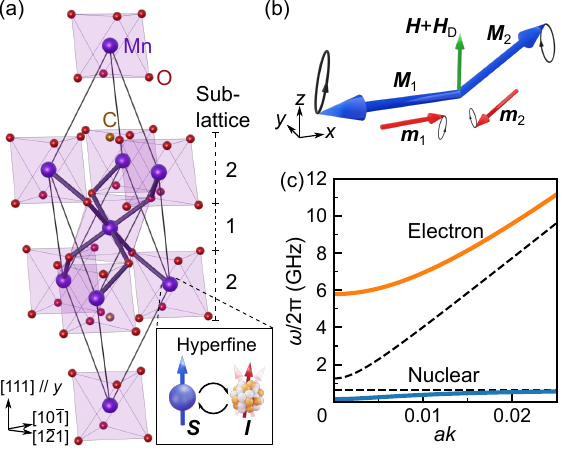}%
 \caption{(a) Crystal structure of MnCO$_3$. (b) Schematic illustration of the hybridized electron and nuclear spin resonance. (c) Dispersion relations of the electron and nuclear spin excitations in MnCO$_3$ [Eqs. (\ref{eq:w0}--\ref{eq:n})]. The colored curves show hybridized dispersions, while the dashed curves show dispersions without hybridization.}
 \label{fig:concept}
\end{figure}

MnCO$_3$ is an insulator whose crystalline structure belongs to the trigonal crystal system \cite{sherman2009electronic, Effenberger1981, Lee2012}.
The rhombohedral unit cell is displayed in Fig. \ref{fig:concept}(a).
A total electron spin $S=5/2$ and a nuclear spin $I=5/2$ (natural isotopic abundance of $^{55}$Mn is 100\%) are located at each Mn site and couple together through the on-site hyperfine interaction.
The electron spins order into the antiferromagnetic state with a finite canting angle due to the Dzyaloshinskii--Moriya (DM) interaction \cite{Dzyaloshinsky1958, Moriya1960}.
The electron spin orientations at different sublattices can be discussed in terms of various internal fields: the exchange field $H_\mathrm{E}$, the DM field $H_\mathrm{D}$, the hard axis anisotropy field $H_\mathrm{K}$ ([111] direction), the easy axis anisotropy field $H_\mathrm{K'}$ ([1$\bar{2}$1] direction) and the hyperfine field from the nuclear spin $H_\mathrm{hf}$ \cite{fink1964nuclear}.
Meanwhile, the nuclear spins are paramagnetically polarized along the electron spin direction by the strong on-site hyperfine field of $\mu_0 H_\mathrm{n} \approx 60$ T, where $\mu_0$ is the vacuum permeability.
The orientations of electron and nuclear magnetic moments, $\mathbf{M}_i \propto -\mathbf{S}_i$ and $\mathbf{m}_i \propto \mathbf{I}_i$, are as shown in Fig. \ref{fig:concept}(b).
The subscripts $i=1,\ 2$ represent the sublattice numbers.
Here, we define the coordinate as $x \parallel [1\bar{2}1]$ (easy axis), $y \parallel [111]$ (hard axis), and $z \parallel [10\bar{1}]$.
We apply an external field $H$ along the $z$ direction.
Large $H_\mathrm{E}$ and small $H_\mathrm{K'}$ orient $\mathbf{M}_i$ along the $x$ axis, and $H+ H_\mathrm{D}$ deflects $\mathbf{M}_i$ toward the $z$ direction.
The canting angle of $\mathbf{M}_i$ from the $x$ axis is $\psi \approx \mathrm{arcsin} (H +H_\mathrm{D})/(2H_\mathrm{E}+H_\mathrm{K'})$ under the approximation of $H_\mathrm{E} \gg$ (Other fields).

Two types of antiferromagnetic electron spin resonance modes can be excited \cite{RezendeBook, rezende2022introduction}.
One is a precession of the net magnetization called the weak-ferromagnetic resonance (in-phase mode) and the other is a precession of the N\'eel vector (out-of-phase mode).
We focus on the weak-ferromagnetic resonance which significantly hybridizes with the nuclear spin excitation due to its smaller resonance gap.
From a simplified model for MnCO$_3$ \cite{fink1964nuclear}, one can predict angular frequencies of uncoupled electron and nuclear spin excitation modes,
\begin{eqnarray}
 \omega_{\mathrm{e0}}^2(k) &\simeq& \omega_H(\omega_H+\omega_\mathrm{D}) +2\omega_\mathrm{E}(\omega_\mathrm{K'}+\omega_\mathrm{hf}) +\Gamma^2(k), \label{eq:w0}\\
 \omega_{\mathrm{n0}} &\simeq& \mu_0\gamma_\mathrm{n}H_\mathrm{n}, \label{eq:n0}
\end{eqnarray}
where $\omega_H = \mu_0\gamma_\mathrm{e}H$, 
$\omega_\mathrm{D} = \mu_0\gamma_\mathrm{e}H_\mathrm{D}$,
$\omega_\mathrm{E} = \mu_0\gamma_\mathrm{e}H_\mathrm{E}$,
$\omega_\mathrm{K'} = \mu_0\gamma_\mathrm{e}H_\mathrm{K'}$,
$\omega_\mathrm{hf} = \mu_0\gamma_\mathrm{e}H_\mathrm{hf}$,
$k$ is the wavenumber of magnon,
$\gamma_\mathrm{e}$ is the gyromagnetic ratio of electron, and
$\gamma_\mathrm{n} = 2\pi \times 10.553$ MHz/T is the gyromagnetic ratio of $^{55}$Mn nucleus in MnCO$_3$ \cite{Mims1967}.
$\Gamma^2(k) = (1-\gamma_k)(\mu_0\gamma_\mathrm{e}H_\mathrm{E})^2$ is a $k$-dependent term with the form factor $\gamma_{k} = \frac{1}{z}\sum_{\boldsymbol{\delta}} e^{i\boldsymbol{\delta}\cdot\mathbf{k}} \simeq 1 - (ak)^2/6$,
the location of nearest neighbors $\boldsymbol{\delta}$, 
and the lattice constant $a=0.4768$ nm\ \cite{Effenberger1981, Lee2012}.
The hybridized electron and nuclear spin excitation modes read
\begin{eqnarray}
 \omega_{\mathrm{e}}^2(k) &\simeq& \omega_\mathrm{e0}^2(k) + G^2, \label{eq:w}\\
 \omega_{\mathrm{n}}^2(k) &\simeq& \omega_\mathrm{n0}^2 - G^2, \label{eq:n}
\end{eqnarray}
where 
\begin{equation}
G = \sqrt{\frac{2\omega_\mathrm{E}\omega_\mathrm{hf}\omega_\mathrm{n0}^2 }{\omega_\mathrm{e0}^2(k) -\omega_\mathrm{n0}^2}} 
\end{equation}
is a detuning parameter due to the hybridization by the transverse dynamical components of $\mathbf{S}$ and $\mathbf{I}$.
The coupling diminishes as $\omega_\mathrm{hf}$ approaches zero ($G\rightarrow 0$).
These uncoupled/hybridized electron and nuclear spin excitation modes, described by Eqs. (\ref{eq:w0})--(\ref{eq:n}), are plotted in Fig. \ref{fig:concept}(c) using the material parameters obtained in this study (as shown later).
For smaller $k$, the uncoupled $\omega_\mathrm{e0}(k)$ approaches $\omega_\mathrm{n0}$, and $\omega_\mathrm{e}(k)$ and $\omega_\mathrm{n}(k)$ repel each other due to the hybridization.

\begin{figure}
 \includegraphics[width=45mm]{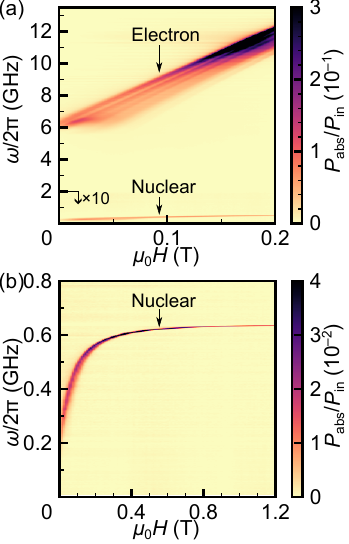}%
 \caption{(a) Observed microwave absorption by MnCO$_3$ at 1.8 K. The displayed $P_\mathrm{abs}$ below 2 GHz is multiplied by 10 for clarity. (b) Observed microwave absorption at 1.8 K in the low-frequency region.}
 \label{fig:maindata}
\end{figure}

We experimentally obtain the hybridized nuclear and electron spin dispersions in MnCO$_3$ using a broadband microwave spectroscopy technique.
A bulk MnCO$_3$ crystal (3 $\times$ 3 $\times$ 0.5 mm$^3$) is set on a coplanar waveguide.
The sample is located in a cryostat with a superconducting magnet to apply $H$ to the $z \parallel$ [10$\bar{1}$] direction.
By using a network analyzer, we irradiate a microwave at the angular frequency $\omega$ and the input power $P_\mathrm{in} =-5$ dBm to the coplanar waveguide to induce a microwave magnetic field parallel to the $x \parallel$ [1$\bar{2}$1] direction and measure the ratio of absorbed and input microwave powers $P_\mathrm{abs}/P_\mathrm{in}$.
We designed two coplanar waveguides.
One is used for most of the measurements and the other improved one is used for the main data in Fig. \ref{fig:maindata}(a).


\begin{figure*}
 \includegraphics[width=130mm]{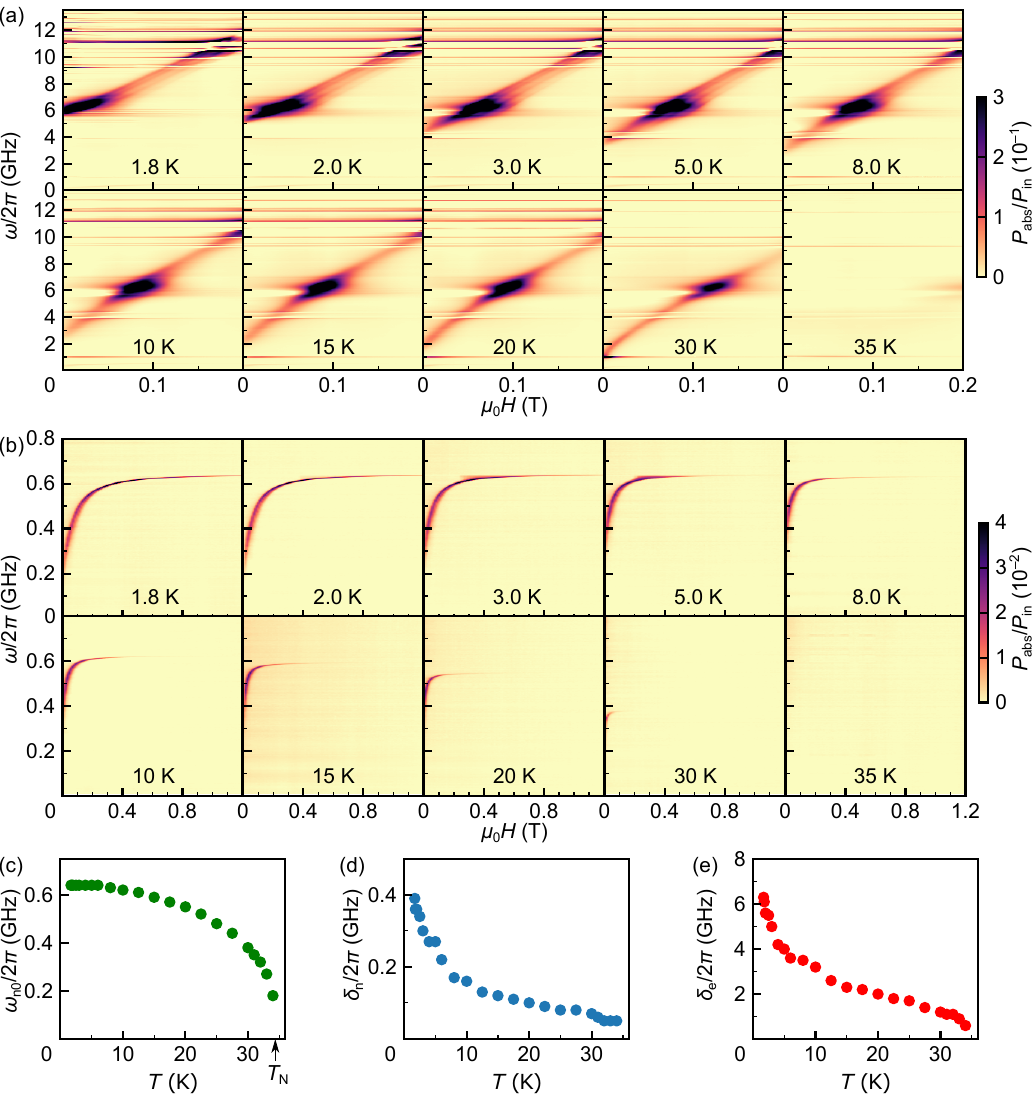}
 \caption{(a) Microwave absorptions at different temperatures for MnCO$_3$. (b) Magnified view of the microwave absorptions in the low-frequency region. (c) Temperature dependence of the nuclear spin excitation frequency at a large magnetic field (1.2 T), i.e. $\omega_\mathrm{n0} \approx \lim_{H\rightarrow \infty}\omega_\mathrm{n}$.  The arrow indicates the N\'eel temperature $T_\mathrm{N}=34.4$ K. (d) Temperature dependence of the pulled frequency of the nuclear branch $\delta_\mathrm{n} \equiv \omega_\mathrm{n0} - \omega_\mathrm{n}(H=0)$. (e) Temperature dependence of the electron spin excitation gap $\delta_\mathrm{e} \equiv \omega_\mathrm{e}(H=0)$.}
 \label{fig:T-dep}
\end{figure*}

Figures \ref{fig:maindata}(a) and (b) show the observed $P_\mathrm{abs}/P_\mathrm{in}$ representing magnetic resonance in MnCO$_3$ at 1.8 K, well below the N\'eel temperature of MnCO$_3$.
The absorption above 6 GHz is due to the hybridized electron spin excitation mode, being almost linearly proportional to the external field $H$.
The multiple lines can be attributed to magnetostatic standing wave modes in the thickness direction of the sample \cite{Beeman1966, Beeman1966b}.
The hybridized nuclear spin excitation mode appears in the low-frequency region in Fig. 2(a).
Figure \ref{fig:maindata}(b) is a magnified view around the low-frequency region of the spectrum shown in Fig. 2(a).
The strong frequency drop near $H= 0$ is the signature of the nuclear spin wave formation known as the frequency pulling effect \cite{deGennes1963, fink1964nuclear}.
The nuclear branch approaches the uncoupled value $\omega_\mathrm{n0}/2\pi \approx$ 640 MHz at large $H$ as the hybridization weakens due to the increasing frequency difference between nuclear and electron spin excitations.
The linewidth is small at large $H$, whereas it broadens at small $H$.
The increase of linewidth at $H\approx 0$ is attributed to the significant hybridization with the dissipative electron spin excitation.
Usually, the nuclear spin excitation is invisible in broadband microwave spectroscopy due to the small nuclear magnetic moment.
The large absorption of the nuclear branch can be attributed to the strong hybridization with a large electron magnetic moment.

We estimate the damping coefficients, coupling constant, and cooperativity for the coupling between nuclear and electron spins from the data in Fig. \ref{fig:maindata}.
The damping coefficients of $\kappa_\mathrm{e}/2\pi \approx 200$ MHz for the electron spin excitation mode and $\kappa_\mathrm{n}/2\pi \approx 1$ MHz for the nuclear spin excitation mode are taken from the asymptotic linewidths at large $H$.
We assumed that the four magnetostatic modes are equally separated and have the same linewidth.
The linewidths at $H\approx 0$ are 200 MHz for the electron spin excitation mode and 100 MHz for the nuclear spin excitation mode, where the broadening of the nuclear branch can be attributed to the hybridization.
We estimate the coupling constant $g$ from a model Hamiltonian $\mathcal{H}/\hbar = \omega_a a^\dag a +\omega_b b^\dag b +g(a^\dag b + a b^\dag)$, where $\omega_a$, $a^\dag$ and $a$ are the uncoupled frequency, creation and annihilation operators for the electron spin excitation, $\omega_b$, $b^\dag$ and $b$ are those for the nuclear spin excitation, and $\hbar$ is the reduced Planck constant.
Diagonalizing $\mathcal{H}$ gives
\begin{equation}
 \omega_\mathrm{e}, \omega_\mathrm{n} = \frac{1}{2}(\omega_a+\omega_b) \pm \frac{1}{2}\sqrt{(\omega_a-\omega_b)^2 +4g^2}.
\end{equation}
Using Eqs. (\ref{eq:w0})--(\ref{eq:n}), $\omega_a = \omega_\mathrm{e0}$, and $\omega_b = \omega_\mathrm{n0}$, we have $g/2\pi = 1.2$ GHz at 1.7 K and 0 T.
As a result, the electron and nuclear spins in MnCO$_3$ are in the strong coupling regime ($g > \kappa_\mathrm{e}, \kappa_\mathrm{n}$) \cite{zhang2014strongly} with the cooperativity $C = g^2/\kappa_\mathrm{e}\kappa_\mathrm{n} \approx 7000$.
This value is greater than those of magnon--magnon ($C\sim 10$) \cite{macneill2019gigahertz, liensberger2019exchange} and magnon--phonon ($C\sim 3000$) \cite{hioki2022coherent} couplings, but is comparable to those of millimeter-sized cavity--magnon systems \cite{huebl2013high, tabuchi2014hybridizing, zhang2014strongly, bourhill2016ultrahigh, lachance2019hybrid, li2019strong, hou2019strong}.

Figures \ref{fig:T-dep}(a) and (b) show the microwave absorption spectra at different temperatures $T$.
The electron branch shifts toward lower frequencies as $T$ increases.
The frequency pulling and absorption intensity of the hybridized nuclear spin excitation mode weaken as $T$ increases.
The dark spot at 6 GHz and horizontal lines are attributed to the effect of microwave standing waves in the coplanar waveguide.

The nuclear spin excitation frequency $\omega_\mathrm{n0}$ in Fig. \ref{fig:T-dep}(c) directly gives the electron magnetic order parameter $\langle S_z \rangle$ through $\omega_\mathrm{n0}=A\langle S_z \rangle /\hbar$, where $A$ is the hyperfine coefficient.
A fitting of $\omega_\mathrm{n0}$ according to $\langle S_z\rangle \propto [1-T/T_\mathrm{N}]^\beta$ gives the N\'eel temperature $T_\mathrm{N} =34.4$ K and the critical exponent $\beta=0.310(3)$ for the antiferromagnetic ordering.
The relatively high $T_\mathrm{N}$ implies a good quality of the sample \cite{Lee2012}.
The critical exponent agrees with values from experiments on weakly anisotropic antiferromagnets \cite{Lee2012, meijer1970some, pincini2018role}.

The pulling frequency $\delta_\mathrm{n} \equiv \omega_\mathrm{n0} - \omega_\mathrm{n}(H=0)$ represents the bandwidth on the nuclear spin wave dispersion.
The temperature dependence of $\delta_\mathrm{n}$, plotted in Fig. \ref{fig:T-dep}(d), represents that the hybridization rapidly increases below $\sim 10$ K, driven by the growing nuclear spin polarization $\langle I_z \rangle \propto 1/T$.
The gap of the electron mode $\delta_\mathrm{e}$ accordingly increases in Fig. \ref{fig:T-dep}(e).
Although the nuclear spin polarization is $\langle I_z \rangle/I \approx $ 2\% at $T=1.8$ K, the pulled frequency of the nuclear branch $\delta_\mathrm{n}/2\pi \equiv [\omega_\mathrm{n0} - \omega_\mathrm{n}(H=0)]/2\pi \approx 400$ MHz is about 60\% of the uncoupled nuclear spin excitation frequency $\omega_\mathrm{n0}/2\pi = $ 640 MHz.
This large shift implies strong hybridization due to the exchange amplification in weakly anisotropic antiferromagnet \cite{turov1966coupled, Shaltiel1966, Andrienko1991, Abdurakhimov2015}.
This effect is described in the detuning parameter $G\propto \sqrt{2\omega_\mathrm{E}\omega_\mathrm{hf}}$ with large $\omega_\mathrm{E}/\pi = 1.87$ THz and $\omega_\mathrm{hf} \propto \langle I_z \rangle$.
The full $T$-dependence data give the internal field values: $\mu_0H_\mathrm{E} =33.4$ T, $\mu_0H_\mathrm{D}= 0.461$ T, $\mu_0H_\mathrm{K'} =0.03$ mT, $\mu_0H_\mathrm{hf} = (1.1\ \mathrm{mT})/(T/\mathrm{K})$, and $\mu_0H_\mathrm{n} =  60.65$ T.

\begin{figure}
 \includegraphics[width=80mm]{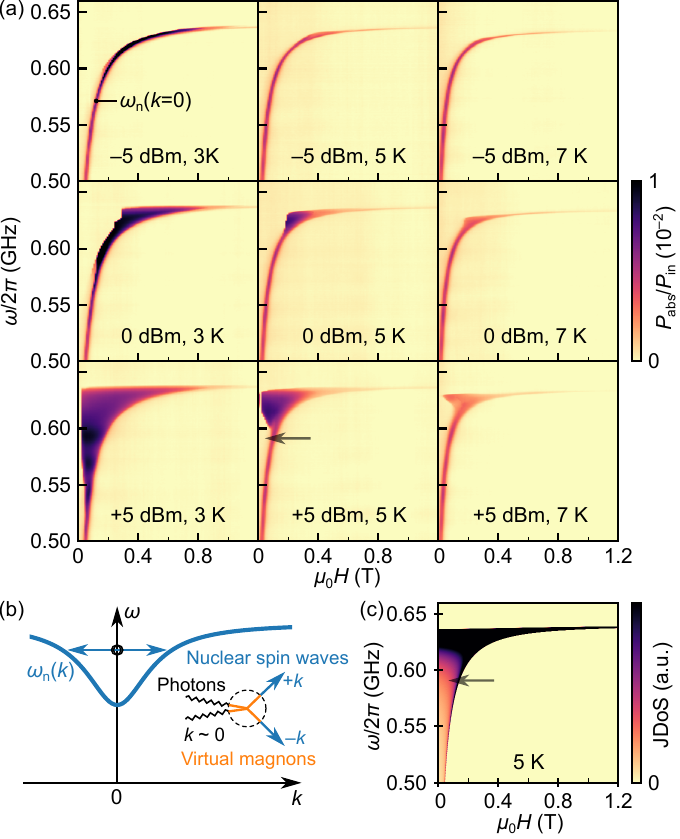}
 \caption{(a) Observed nonlinear microwave absorption at different input microwave powers and temperatures. The arrow indicates the zero-absorption pocket (5 dBm, 5K). (b) Dispersion relation of the nuclear spin wave. The diagram shows a process for generating the finite wavenumber nuclear spin wave. (c) Calculated joint density of state. The arrow indicates the absence of JDoS.}
 \label{fig:nonlinear}
\end{figure}

We found a nonlinear regime of the hybridized nuclear spin excitation mode under the large input microwave power $P_\mathrm{in}$.
Figure \ref{fig:nonlinear}(a) depicts the microwave absorption at different $P_\mathrm{in}$ and $T$.
With $P_\mathrm{in} =-5$ dBm and less, the nuclear branch obeys $\omega_\mathrm{n}(k=0)$ in Eq. (\ref{eq:n}).
By contrast, with $P_\mathrm{in} = 0$ and $+5$ dBm, the absorption spread toward higher frequencies.
The spread appears to be bounded by the upper limit at $\omega_\mathrm{n0}/2\pi$ = 640 MHz and the lower limit of $\omega_\mathrm{n}(k=0)$.
The nonlinear absorption is prominent at lower $T$'s, at which the hybridization is strong.
Note that we swept the frequency from 0.66 to 0.50 GHz with a long dwell time to prevent history effects.

Several nonlinear responses of hybridized nuclear spin excitation mode in MnCO$_3$ have already been reported \cite{fink1964nuclear, Shaltiel1966, Abdurakhimov2018}.
One possible explanation is an increase of the nuclear spin temperature $T_\mathrm{n}$ from the lattice and electron temperature $T$ \cite{fink1964nuclear, Shaltiel1966}.
In this case, the absorption shifts toward a higher frequency.
However, the absorptions in Fig. \ref{fig:nonlinear}(a) keep the low-frequency edge at $\omega_n(k=0,\ T)$, meaning that $T_\mathrm{n}\approx T$.
Another possibility is Duffing (Kerr) nonlinearity of the uniform mode ($k=0$) \cite{Abdurakhimov2018}.
In this case, the absorption shows a single triangular peak in the frequency sweep.
However, the absence of the absorption (zero-absorption pocket) in the middle-frequency region [indicated by the arrow in Fig. \ref{fig:nonlinear}(a) (5 dBm, 5 K)] contradicts with this single-peak scenario.

We attribute the observed nonlinear absorption to the scattering into finite wavenumber nuclear spin waves.
A process to create finite wavenumber nuclear magnons is allowed through virtual electron magnon scatterings, described in Fig. \ref{fig:nonlinear}(b).
The rate of the process is proportional to the joint density of state (JDoS) of the hybridized nuclear spin wave dispersion,
\begin{equation}
 J(\omega) = \sum_\mathbf{k} \delta(\omega_{\mathrm{n}}(k)-\omega).
 \label{eq:jdos}
\end{equation}
Figure \ref{fig:nonlinear}(c) shows calculated JDoS, having a low-frequency limit at $\omega_\mathrm{n}(k=0)$, a spread dense area at high frequencies, and an upper limit at $\omega_\mathrm{n0}$.
The location of the JDoS absence indicated by the arrow in Fig. \ref{fig:nonlinear}(c) matches well with that of the zero-absorption pocket, being consistent with the scenario shown in Fig. \ref{fig:nonlinear}(b).

In conclusion, we measured the field dispersion of the hybridized nuclear and electron spin excitation in MnCO$_3$ by using the broadband microwave spectroscopy technique.
We quantified the internal hyperfine coupling fields, the cooperativity, the critical exponent, and the N\'eel temperature from the observed dispersions.
The nonlinearity in the hybridized nuclear spin excitation mode implies the generation of the finite wavenumber nuclear spin waves with relatively small microwave fields.
The high cooperativity in MnCO$_3$ may be useful for applications such as electric control of nuclear spin dynamics.

\begin{acknowledgments}
 We thank S. Daimon, T. Sugimoto, and T. Hioki for discussions and technical support.
 This work was partially supported by JSPS KAKENHI (Nos. JP19H05600, JP20H02599, JP20K15160, JP22H05114, JP23KJ0678, and JP24K01326), JST CREST (Nos. JPMJCR20C1 and JPMJCR20T2), MEXT Initiative to Establish Next-generation Novel Integrated Circuits Centers (X-NICS) (No. JPJ011438), Murata Science Foundation, and Institute for AI and Beyond of the University of Tokyo.
\end{acknowledgments}


\begin{thebibliography}{37}%
\makeatletter
\providecommand \@ifxundefined [1]{%
 \@ifx{#1\undefined}
}%
\providecommand \@ifnum [1]{%
 \ifnum #1\expandafter \@firstoftwo
 \else \expandafter \@secondoftwo
 \fi
}%
\providecommand \@ifx [1]{%
 \ifx #1\expandafter \@firstoftwo
 \else \expandafter \@secondoftwo
 \fi
}%
\providecommand \natexlab [1]{#1}%
\providecommand \enquote  [1]{``#1''}%
\providecommand \bibnamefont  [1]{#1}%
\providecommand \bibfnamefont [1]{#1}%
\providecommand \citenamefont [1]{#1}%
\providecommand \href@noop [0]{\@secondoftwo}%
\providecommand \href [0]{\begingroup \@sanitize@url \@href}%
\providecommand \@href[1]{\@@startlink{#1}\@@href}%
\providecommand \@@href[1]{\endgroup#1\@@endlink}%
\providecommand \@sanitize@url [0]{\catcode `\\12\catcode `\$12\catcode
  `\&12\catcode `\#12\catcode `\^12\catcode `\_12\catcode `\%12\relax}%
\providecommand \@@startlink[1]{}%
\providecommand \@@endlink[0]{}%
\providecommand \url  [0]{\begingroup\@sanitize@url \@url }%
\providecommand \@url [1]{\endgroup\@href {#1}{\urlprefix }}%
\providecommand \urlprefix  [0]{URL }%
\providecommand \Eprint [0]{\href }%
\providecommand \doibase [0]{https://doi.org/}%
\providecommand \selectlanguage [0]{\@gobble}%
\providecommand \bibinfo  [0]{\@secondoftwo}%
\providecommand \bibfield  [0]{\@secondoftwo}%
\providecommand \translation [1]{[#1]}%
\providecommand \BibitemOpen [0]{}%
\providecommand \bibitemStop [0]{}%
\providecommand \bibitemNoStop [0]{.\EOS\space}%
\providecommand \EOS [0]{\spacefactor3000\relax}%
\providecommand \BibitemShut  [1]{\csname bibitem#1\endcsname}%
\let\auto@bib@innerbib\@empty
\bibitem [{\citenamefont {Abragam}(1961)}]{abragam1961principles}%
  \BibitemOpen
  \bibfield  {author} {\bibinfo {author} {\bibfnamefont {A.}~\bibnamefont
  {Abragam}},\ }\href@noop {} {\emph {\bibinfo {title} {The principles of
  nuclear magnetism}}},\ \bibinfo {number} {32}\ (\bibinfo  {publisher} {Oxford
  university press},\ \bibinfo {year} {1961})\BibitemShut {NoStop}%
\bibitem [{\citenamefont {Shiomi}\ \emph {et~al.}(2019)\citenamefont {Shiomi},
  \citenamefont {Lustikova}, \citenamefont {Watanabe}, \citenamefont {Hirobe},
  \citenamefont {Takahashi},\ and\ \citenamefont {Saitoh}}]{Shiomi2019}%
  \BibitemOpen
  \bibfield  {author} {\bibinfo {author} {\bibfnamefont {Y.}~\bibnamefont
  {Shiomi}}, \bibinfo {author} {\bibfnamefont {J.}~\bibnamefont {Lustikova}},
  \bibinfo {author} {\bibfnamefont {S.}~\bibnamefont {Watanabe}}, \bibinfo
  {author} {\bibfnamefont {D.}~\bibnamefont {Hirobe}}, \bibinfo {author}
  {\bibfnamefont {S.}~\bibnamefont {Takahashi}},\ and\ \bibinfo {author}
  {\bibfnamefont {E.}~\bibnamefont {Saitoh}},\ }\href@noop {} {\bibfield
  {journal} {\bibinfo  {journal} {Nature Physics}\ }\textbf {\bibinfo {volume}
  {15}},\ \bibinfo {pages} {22} (\bibinfo {year} {2019})}\BibitemShut {NoStop}%
\bibitem [{\citenamefont {Kikkawa}\ \emph {et~al.}(2021)\citenamefont
  {Kikkawa}, \citenamefont {Reitz}, \citenamefont {Ito}, \citenamefont
  {Makiuchi}, \citenamefont {Sugimoto}, \citenamefont {Tsunekawa},
  \citenamefont {Daimon}, \citenamefont {Oyanagi}, \citenamefont {Ramos},
  \citenamefont {Takahashi}, \citenamefont {Shiomi}, \citenamefont
  {Tserkovnyak},\ and\ \citenamefont {Saitoh}}]{Kikkawa2021}%
  \BibitemOpen
  \bibfield  {author} {\bibinfo {author} {\bibfnamefont {T.}~\bibnamefont
  {Kikkawa}}, \bibinfo {author} {\bibfnamefont {D.}~\bibnamefont {Reitz}},
  \bibinfo {author} {\bibfnamefont {H.}~\bibnamefont {Ito}}, \bibinfo {author}
  {\bibfnamefont {T.}~\bibnamefont {Makiuchi}}, \bibinfo {author}
  {\bibfnamefont {T.}~\bibnamefont {Sugimoto}}, \bibinfo {author}
  {\bibfnamefont {K.}~\bibnamefont {Tsunekawa}}, \bibinfo {author}
  {\bibfnamefont {S.}~\bibnamefont {Daimon}}, \bibinfo {author} {\bibfnamefont
  {K.}~\bibnamefont {Oyanagi}}, \bibinfo {author} {\bibfnamefont
  {R.}~\bibnamefont {Ramos}}, \bibinfo {author} {\bibfnamefont
  {S.}~\bibnamefont {Takahashi}}, \bibinfo {author} {\bibfnamefont
  {Y.}~\bibnamefont {Shiomi}}, \bibinfo {author} {\bibfnamefont
  {Y.}~\bibnamefont {Tserkovnyak}},\ and\ \bibinfo {author} {\bibfnamefont
  {E.}~\bibnamefont {Saitoh}},\ }\href@noop {} {\bibfield  {journal} {\bibinfo
  {journal} {Nature Communications}\ }\textbf {\bibinfo {volume} {12}},\
  \bibinfo {pages} {4356} (\bibinfo {year} {2021})}\BibitemShut {NoStop}%
\bibitem [{\citenamefont {Rezende}(2022)}]{rezende2022introduction}%
  \BibitemOpen
  \bibfield  {author} {\bibinfo {author} {\bibfnamefont {S.~M.}\ \bibnamefont
  {Rezende}},\ }\href@noop {} {\bibfield  {journal} {\bibinfo  {journal}
  {Journal of Applied Physics}\ }\textbf {\bibinfo {volume} {132}},\ \bibinfo
  {pages} {091101} (\bibinfo {year} {2022})}\BibitemShut {NoStop}%
\bibitem [{\citenamefont {Maekawa}\ \emph {et~al.}(2023)\citenamefont
  {Maekawa}, \citenamefont {Kikkawa}, \citenamefont {Chudo}, \citenamefont
  {Ieda},\ and\ \citenamefont {Saitoh}}]{maekawa2023spin}%
  \BibitemOpen
  \bibfield  {author} {\bibinfo {author} {\bibfnamefont {S.}~\bibnamefont
  {Maekawa}}, \bibinfo {author} {\bibfnamefont {T.}~\bibnamefont {Kikkawa}},
  \bibinfo {author} {\bibfnamefont {H.}~\bibnamefont {Chudo}}, \bibinfo
  {author} {\bibfnamefont {J.}~\bibnamefont {Ieda}},\ and\ \bibinfo {author}
  {\bibfnamefont {E.}~\bibnamefont {Saitoh}},\ }\href@noop {} {\bibfield
  {journal} {\bibinfo  {journal} {Journal of Applied Physics}\ }\textbf
  {\bibinfo {volume} {133}},\ \bibinfo {pages} {020902} (\bibinfo {year}
  {2023})}\BibitemShut {NoStop}%
\bibitem [{\citenamefont {Kikkawa}\ and\ \citenamefont
  {Saitoh}(2023)}]{kikkawa2023spin}%
  \BibitemOpen
  \bibfield  {author} {\bibinfo {author} {\bibfnamefont {T.}~\bibnamefont
  {Kikkawa}}\ and\ \bibinfo {author} {\bibfnamefont {E.}~\bibnamefont
  {Saitoh}},\ }\href@noop {} {\bibfield  {journal} {\bibinfo  {journal} {Annual
  Review of Condensed Matter Physics}\ }\textbf {\bibinfo {volume} {14}},\
  \bibinfo {pages} {129} (\bibinfo {year} {2023})}\BibitemShut {NoStop}%
\bibitem [{\citenamefont {Chigusa}\ \emph {et~al.}(2023)\citenamefont
  {Chigusa}, \citenamefont {Moroi}, \citenamefont {Nakayama},\ and\
  \citenamefont {Sichanugrist}}]{chigusa2023dark}%
  \BibitemOpen
  \bibfield  {author} {\bibinfo {author} {\bibfnamefont {S.}~\bibnamefont
  {Chigusa}}, \bibinfo {author} {\bibfnamefont {T.}~\bibnamefont {Moroi}},
  \bibinfo {author} {\bibfnamefont {K.}~\bibnamefont {Nakayama}},\ and\
  \bibinfo {author} {\bibfnamefont {T.}~\bibnamefont {Sichanugrist}},\ }\href
  {https://doi.org/10.1103/PhysRevD.108.095007} {\bibfield  {journal} {\bibinfo
   {journal} {Phys. Rev. D}\ }\textbf {\bibinfo {volume} {108}},\ \bibinfo
  {pages} {095007} (\bibinfo {year} {2023})}\BibitemShut {NoStop}%
\bibitem [{\citenamefont {Suhl}(1958)}]{Suhl1958}%
  \BibitemOpen
  \bibfield  {author} {\bibinfo {author} {\bibfnamefont {H.}~\bibnamefont
  {Suhl}},\ }\href@noop {} {\bibfield  {journal} {\bibinfo  {journal} {Phys.
  Rev.}\ }\textbf {\bibinfo {volume} {109}},\ \bibinfo {pages} {606} (\bibinfo
  {year} {1958})}\BibitemShut {NoStop}%
\bibitem [{\citenamefont {Nakamura}(1958)}]{Nakamura1958}%
  \BibitemOpen
  \bibfield  {author} {\bibinfo {author} {\bibfnamefont {T.}~\bibnamefont
  {Nakamura}},\ }\href@noop {} {\bibfield  {journal} {\bibinfo  {journal}
  {Prog. Theor. Phys.}\ }\textbf {\bibinfo {volume} {20}},\ \bibinfo {pages}
  {542} (\bibinfo {year} {1958})}\BibitemShut {NoStop}%
\bibitem [{\citenamefont {{de Gennes}}\ \emph {et~al.}(1963)\citenamefont {{de
  Gennes}}, \citenamefont {Pincus}, \citenamefont {Hartmann-Boutron},\ and\
  \citenamefont {Winter}}]{deGennes1963}%
  \BibitemOpen
  \bibfield  {author} {\bibinfo {author} {\bibfnamefont {P.~G.}\ \bibnamefont
  {{de Gennes}}}, \bibinfo {author} {\bibfnamefont {P.~A.}\ \bibnamefont
  {Pincus}}, \bibinfo {author} {\bibfnamefont {F.}~\bibnamefont
  {Hartmann-Boutron}},\ and\ \bibinfo {author} {\bibfnamefont {J.~M.}\
  \bibnamefont {Winter}},\ }\href@noop {} {\bibfield  {journal} {\bibinfo
  {journal} {Phys. Rev.}\ }\textbf {\bibinfo {volume} {129}},\ \bibinfo {pages}
  {1105} (\bibinfo {year} {1963})}\BibitemShut {NoStop}%
\bibitem [{\citenamefont {Andrienko}\ \emph {et~al.}(1991)\citenamefont
  {Andrienko}, \citenamefont {Ozhogin}, \citenamefont {Safonov},\ and\
  \citenamefont {Yakubovskii}}]{Andrienko1991}%
  \BibitemOpen
  \bibfield  {author} {\bibinfo {author} {\bibfnamefont {A.~V.}\ \bibnamefont
  {Andrienko}}, \bibinfo {author} {\bibfnamefont {V.~I.}\ \bibnamefont
  {Ozhogin}}, \bibinfo {author} {\bibfnamefont {V.~L.}\ \bibnamefont
  {Safonov}},\ and\ \bibinfo {author} {\bibfnamefont {A.~Y.}\ \bibnamefont
  {Yakubovskii}},\ }\href@noop {} {\bibfield  {journal} {\bibinfo  {journal}
  {Sov. Phys. Usp.}\ }\textbf {\bibinfo {volume} {34}},\ \bibinfo {pages} {843}
  (\bibinfo {year} {1991})}\BibitemShut {NoStop}%
\bibitem [{\citenamefont {Sherman}(2009)}]{sherman2009electronic}%
  \BibitemOpen
  \bibfield  {author} {\bibinfo {author} {\bibfnamefont {D.~M.}\ \bibnamefont
  {Sherman}},\ }\href@noop {} {\bibfield  {journal} {\bibinfo  {journal}
  {American Mineralogist}\ }\textbf {\bibinfo {volume} {94}},\ \bibinfo {pages}
  {166} (\bibinfo {year} {2009})}\BibitemShut {NoStop}%
\bibitem [{\citenamefont {Effenberger}\ \emph {et~al.}(1981)\citenamefont
  {Effenberger}, \citenamefont {Mereiter},\ and\ \citenamefont
  {Zemann}}]{Effenberger1981}%
  \BibitemOpen
  \bibfield  {author} {\bibinfo {author} {\bibfnamefont {H.}~\bibnamefont
  {Effenberger}}, \bibinfo {author} {\bibfnamefont {K.}~\bibnamefont
  {Mereiter}},\ and\ \bibinfo {author} {\bibfnamefont {J.}~\bibnamefont
  {Zemann}},\ }\href@noop {} {\bibfield  {journal} {\bibinfo  {journal}
  {Zeitschrift f{\"u}r Kristallographie-Crystalline Materials}\ }\textbf
  {\bibinfo {volume} {156}},\ \bibinfo {pages} {233} (\bibinfo {year}
  {1981})}\BibitemShut {NoStop}%
\bibitem [{\citenamefont {Lee}\ \emph {et~al.}(2012)\citenamefont {Lee},
  \citenamefont {Hong}, \citenamefont {Kim}, \citenamefont
  {Jagli{\v{c}}i{\'c}}, \citenamefont {Jazbec}, \citenamefont {Wencka},
  \citenamefont {Jelen},\ and\ \citenamefont {Dolin{\v{s}}ek}}]{Lee2012}%
  \BibitemOpen
  \bibfield  {author} {\bibinfo {author} {\bibfnamefont {J.~B.}\ \bibnamefont
  {Lee}}, \bibinfo {author} {\bibfnamefont {W.~G.}\ \bibnamefont {Hong}},
  \bibinfo {author} {\bibfnamefont {H.~J.}\ \bibnamefont {Kim}}, \bibinfo
  {author} {\bibfnamefont {Z.}~\bibnamefont {Jagli{\v{c}}i{\'c}}}, \bibinfo
  {author} {\bibfnamefont {S.}~\bibnamefont {Jazbec}}, \bibinfo {author}
  {\bibfnamefont {M.}~\bibnamefont {Wencka}}, \bibinfo {author} {\bibfnamefont
  {A.}~\bibnamefont {Jelen}},\ and\ \bibinfo {author} {\bibfnamefont
  {J.}~\bibnamefont {Dolin{\v{s}}ek}},\ }\href@noop {} {\bibfield  {journal}
  {\bibinfo  {journal} {Physical Review B}\ }\textbf {\bibinfo {volume} {86}},\
  \bibinfo {pages} {224407} (\bibinfo {year} {2012})}\BibitemShut {NoStop}%
\bibitem [{\citenamefont {Dzyaloshinsky}(1958)}]{Dzyaloshinsky1958}%
  \BibitemOpen
  \bibfield  {author} {\bibinfo {author} {\bibfnamefont {I.}~\bibnamefont
  {Dzyaloshinsky}},\ }\href@noop {} {\bibfield  {journal} {\bibinfo  {journal}
  {Journal of Physics and Chemistry of Solids}\ }\textbf {\bibinfo {volume}
  {4}},\ \bibinfo {pages} {241} (\bibinfo {year} {1958})}\BibitemShut {NoStop}%
\bibitem [{\citenamefont {Moriya}(1960)}]{Moriya1960}%
  \BibitemOpen
  \bibfield  {author} {\bibinfo {author} {\bibfnamefont {T.}~\bibnamefont
  {Moriya}},\ }\href@noop {} {\bibfield  {journal} {\bibinfo  {journal}
  {Physical Review}\ }\textbf {\bibinfo {volume} {120}},\ \bibinfo {pages} {91}
  (\bibinfo {year} {1960})}\BibitemShut {NoStop}%
\bibitem [{\citenamefont {Fink}\ and\ \citenamefont
  {Shaltiel}(1964)}]{fink1964nuclear}%
  \BibitemOpen
  \bibfield  {author} {\bibinfo {author} {\bibfnamefont {H.}~\bibnamefont
  {Fink}}\ and\ \bibinfo {author} {\bibfnamefont {D.}~\bibnamefont
  {Shaltiel}},\ }\href@noop {} {\bibfield  {journal} {\bibinfo  {journal}
  {Phys. Rev.}\ }\textbf {\bibinfo {volume} {136}},\ \bibinfo {pages} {218}
  (\bibinfo {year} {1964})}\BibitemShut {NoStop}%
\bibitem [{\citenamefont {Rezende}(2020)}]{RezendeBook}%
  \BibitemOpen
  \bibfield  {author} {\bibinfo {author} {\bibfnamefont {S.~M.}\ \bibnamefont
  {Rezende}},\ }\href@noop {} {\emph {\bibinfo {title} {Fundamentals of
  Magnonics}}},\ \bibinfo {edition} {1st}\ ed.\ (\bibinfo  {publisher}
  {Springer},\ \bibinfo {year} {2020})\BibitemShut {NoStop}%
\bibitem [{\citenamefont {Mims}\ \emph {et~al.}(1967)\citenamefont {Mims},
  \citenamefont {Devlin}, \citenamefont {Geschwind},\ and\ \citenamefont
  {Jaccarino}}]{Mims1967}%
  \BibitemOpen
  \bibfield  {author} {\bibinfo {author} {\bibfnamefont {W.~B.}\ \bibnamefont
  {Mims}}, \bibinfo {author} {\bibfnamefont {G.~E.}\ \bibnamefont {Devlin}},
  \bibinfo {author} {\bibfnamefont {S.}~\bibnamefont {Geschwind}},\ and\
  \bibinfo {author} {\bibfnamefont {V.}~\bibnamefont {Jaccarino}},\ }\href@noop
  {} {\bibfield  {journal} {\bibinfo  {journal} {Phyisics Letters}\ }\textbf
  {\bibinfo {volume} {24}},\ \bibinfo {pages} {481} (\bibinfo {year}
  {1967})}\BibitemShut {NoStop}%
\bibitem [{\citenamefont {Beeman}(1966)}]{Beeman1966}%
  \BibitemOpen
  \bibfield  {author} {\bibinfo {author} {\bibfnamefont {D.~E.}\ \bibnamefont
  {Beeman}},\ }\href@noop {} {\bibfield  {journal} {\bibinfo  {journal}
  {Journal of Applied Physics}\ }\textbf {\bibinfo {volume} {37}},\ \bibinfo
  {pages} {1136} (\bibinfo {year} {1966})}\BibitemShut {NoStop}%
\bibitem [{\citenamefont {Beeman}\ \emph {et~al.}(1966)\citenamefont {Beeman},
  \citenamefont {Fink},\ and\ \citenamefont {Shaltiel}}]{Beeman1966b}%
  \BibitemOpen
  \bibfield  {author} {\bibinfo {author} {\bibfnamefont {D.~E.}\ \bibnamefont
  {Beeman}}, \bibinfo {author} {\bibfnamefont {H.~J.}\ \bibnamefont {Fink}},\
  and\ \bibinfo {author} {\bibfnamefont {D.}~\bibnamefont {Shaltiel}},\
  }\href@noop {} {\bibfield  {journal} {\bibinfo  {journal} {Physical Review}\
  }\textbf {\bibinfo {volume} {147}},\ \bibinfo {pages} {454} (\bibinfo {year}
  {1966})}\BibitemShut {NoStop}%
\bibitem [{\citenamefont {Zhang}\ \emph {et~al.}(2014)\citenamefont {Zhang},
  \citenamefont {Zou}, \citenamefont {Jiang},\ and\ \citenamefont
  {Tang}}]{zhang2014strongly}%
  \BibitemOpen
  \bibfield  {author} {\bibinfo {author} {\bibfnamefont {X.}~\bibnamefont
  {Zhang}}, \bibinfo {author} {\bibfnamefont {C.-L.}\ \bibnamefont {Zou}},
  \bibinfo {author} {\bibfnamefont {L.}~\bibnamefont {Jiang}},\ and\ \bibinfo
  {author} {\bibfnamefont {H.~X.}\ \bibnamefont {Tang}},\ }\href
  {https://doi.org/10.1103/PhysRevLett.113.156401} {\bibfield  {journal}
  {\bibinfo  {journal} {Phys. Rev. Lett.}\ }\textbf {\bibinfo {volume} {113}},\
  \bibinfo {pages} {156401} (\bibinfo {year} {2014})}\BibitemShut {NoStop}%
\bibitem [{\citenamefont {MacNeill}\ \emph {et~al.}(2019)\citenamefont
  {MacNeill}, \citenamefont {Hou}, \citenamefont {Klein}, \citenamefont
  {Zhang}, \citenamefont {Jarillo-Herrero},\ and\ \citenamefont
  {Liu}}]{macneill2019gigahertz}%
  \BibitemOpen
  \bibfield  {author} {\bibinfo {author} {\bibfnamefont {D.}~\bibnamefont
  {MacNeill}}, \bibinfo {author} {\bibfnamefont {J.~T.}\ \bibnamefont {Hou}},
  \bibinfo {author} {\bibfnamefont {D.~R.}\ \bibnamefont {Klein}}, \bibinfo
  {author} {\bibfnamefont {P.}~\bibnamefont {Zhang}}, \bibinfo {author}
  {\bibfnamefont {P.}~\bibnamefont {Jarillo-Herrero}},\ and\ \bibinfo {author}
  {\bibfnamefont {L.}~\bibnamefont {Liu}},\ }\href@noop {} {\bibfield
  {journal} {\bibinfo  {journal} {Physical Review Letters}\ }\textbf {\bibinfo
  {volume} {123}},\ \bibinfo {pages} {047204} (\bibinfo {year}
  {2019})}\BibitemShut {NoStop}%
\bibitem [{\citenamefont {Liensberger}\ \emph {et~al.}(2019)\citenamefont
  {Liensberger}, \citenamefont {Kamra}, \citenamefont {Maier-Flaig},
  \citenamefont {Gepr{\"a}gs}, \citenamefont {Erb}, \citenamefont
  {Goennenwein}, \citenamefont {Gross}, \citenamefont {Belzig}, \citenamefont
  {Huebl},\ and\ \citenamefont {Weiler}}]{liensberger2019exchange}%
  \BibitemOpen
  \bibfield  {author} {\bibinfo {author} {\bibfnamefont {L.}~\bibnamefont
  {Liensberger}}, \bibinfo {author} {\bibfnamefont {A.}~\bibnamefont {Kamra}},
  \bibinfo {author} {\bibfnamefont {H.}~\bibnamefont {Maier-Flaig}}, \bibinfo
  {author} {\bibfnamefont {S.}~\bibnamefont {Gepr{\"a}gs}}, \bibinfo {author}
  {\bibfnamefont {A.}~\bibnamefont {Erb}}, \bibinfo {author} {\bibfnamefont
  {S.~T.}\ \bibnamefont {Goennenwein}}, \bibinfo {author} {\bibfnamefont
  {R.}~\bibnamefont {Gross}}, \bibinfo {author} {\bibfnamefont
  {W.}~\bibnamefont {Belzig}}, \bibinfo {author} {\bibfnamefont
  {H.}~\bibnamefont {Huebl}},\ and\ \bibinfo {author} {\bibfnamefont
  {M.}~\bibnamefont {Weiler}},\ }\href@noop {} {\bibfield  {journal} {\bibinfo
  {journal} {Physical Review Letters}\ }\textbf {\bibinfo {volume} {123}},\
  \bibinfo {pages} {117204} (\bibinfo {year} {2019})}\BibitemShut {NoStop}%
\bibitem [{\citenamefont {Hioki}\ \emph {et~al.}(2022)\citenamefont {Hioki},
  \citenamefont {Hashimoto},\ and\ \citenamefont {Saitoh}}]{hioki2022coherent}%
  \BibitemOpen
  \bibfield  {author} {\bibinfo {author} {\bibfnamefont {T.}~\bibnamefont
  {Hioki}}, \bibinfo {author} {\bibfnamefont {Y.}~\bibnamefont {Hashimoto}},\
  and\ \bibinfo {author} {\bibfnamefont {E.}~\bibnamefont {Saitoh}},\
  }\href@noop {} {\bibfield  {journal} {\bibinfo  {journal} {Communications
  Physics}\ }\textbf {\bibinfo {volume} {5}},\ \bibinfo {pages} {115} (\bibinfo
  {year} {2022})}\BibitemShut {NoStop}%
\bibitem [{\citenamefont {Huebl}\ \emph {et~al.}(2013)\citenamefont {Huebl},
  \citenamefont {Zollitsch}, \citenamefont {Lotze}, \citenamefont {Hocke},
  \citenamefont {Greifenstein}, \citenamefont {Marx}, \citenamefont {Gross},\
  and\ \citenamefont {Goennenwein}}]{huebl2013high}%
  \BibitemOpen
  \bibfield  {author} {\bibinfo {author} {\bibfnamefont {H.}~\bibnamefont
  {Huebl}}, \bibinfo {author} {\bibfnamefont {C.~W.}\ \bibnamefont
  {Zollitsch}}, \bibinfo {author} {\bibfnamefont {J.}~\bibnamefont {Lotze}},
  \bibinfo {author} {\bibfnamefont {F.}~\bibnamefont {Hocke}}, \bibinfo
  {author} {\bibfnamefont {M.}~\bibnamefont {Greifenstein}}, \bibinfo {author}
  {\bibfnamefont {A.}~\bibnamefont {Marx}}, \bibinfo {author} {\bibfnamefont
  {R.}~\bibnamefont {Gross}},\ and\ \bibinfo {author} {\bibfnamefont {S.~T.}\
  \bibnamefont {Goennenwein}},\ }\href@noop {} {\bibfield  {journal} {\bibinfo
  {journal} {Physical Review Letters}\ }\textbf {\bibinfo {volume} {111}},\
  \bibinfo {pages} {127003} (\bibinfo {year} {2013})}\BibitemShut {NoStop}%
\bibitem [{\citenamefont {Tabuchi}\ \emph {et~al.}(2014)\citenamefont
  {Tabuchi}, \citenamefont {Ishino}, \citenamefont {Ishikawa}, \citenamefont
  {Yamazaki}, \citenamefont {Usami},\ and\ \citenamefont
  {Nakamura}}]{tabuchi2014hybridizing}%
  \BibitemOpen
  \bibfield  {author} {\bibinfo {author} {\bibfnamefont {Y.}~\bibnamefont
  {Tabuchi}}, \bibinfo {author} {\bibfnamefont {S.}~\bibnamefont {Ishino}},
  \bibinfo {author} {\bibfnamefont {T.}~\bibnamefont {Ishikawa}}, \bibinfo
  {author} {\bibfnamefont {R.}~\bibnamefont {Yamazaki}}, \bibinfo {author}
  {\bibfnamefont {K.}~\bibnamefont {Usami}},\ and\ \bibinfo {author}
  {\bibfnamefont {Y.}~\bibnamefont {Nakamura}},\ }\href@noop {} {\bibfield
  {journal} {\bibinfo  {journal} {Physical Review Letters}\ }\textbf {\bibinfo
  {volume} {113}},\ \bibinfo {pages} {083603} (\bibinfo {year}
  {2014})}\BibitemShut {NoStop}%
\bibitem [{\citenamefont {Bourhill}\ \emph {et~al.}(2016)\citenamefont
  {Bourhill}, \citenamefont {Kostylev}, \citenamefont {Goryachev},
  \citenamefont {Creedon},\ and\ \citenamefont
  {Tobar}}]{bourhill2016ultrahigh}%
  \BibitemOpen
  \bibfield  {author} {\bibinfo {author} {\bibfnamefont {J.}~\bibnamefont
  {Bourhill}}, \bibinfo {author} {\bibfnamefont {N.}~\bibnamefont {Kostylev}},
  \bibinfo {author} {\bibfnamefont {M.}~\bibnamefont {Goryachev}}, \bibinfo
  {author} {\bibfnamefont {D.}~\bibnamefont {Creedon}},\ and\ \bibinfo {author}
  {\bibfnamefont {M.}~\bibnamefont {Tobar}},\ }\href@noop {} {\bibfield
  {journal} {\bibinfo  {journal} {Physical Review B}\ }\textbf {\bibinfo
  {volume} {93}},\ \bibinfo {pages} {144420} (\bibinfo {year}
  {2016})}\BibitemShut {NoStop}%
\bibitem [{\citenamefont {Lachance-Quirion}\ \emph {et~al.}(2019)\citenamefont
  {Lachance-Quirion}, \citenamefont {Tabuchi}, \citenamefont {Gloppe},
  \citenamefont {Usami},\ and\ \citenamefont {Nakamura}}]{lachance2019hybrid}%
  \BibitemOpen
  \bibfield  {author} {\bibinfo {author} {\bibfnamefont {D.}~\bibnamefont
  {Lachance-Quirion}}, \bibinfo {author} {\bibfnamefont {Y.}~\bibnamefont
  {Tabuchi}}, \bibinfo {author} {\bibfnamefont {A.}~\bibnamefont {Gloppe}},
  \bibinfo {author} {\bibfnamefont {K.}~\bibnamefont {Usami}},\ and\ \bibinfo
  {author} {\bibfnamefont {Y.}~\bibnamefont {Nakamura}},\ }\href@noop {}
  {\bibfield  {journal} {\bibinfo  {journal} {Applied Physics Express}\
  }\textbf {\bibinfo {volume} {12}},\ \bibinfo {pages} {070101} (\bibinfo
  {year} {2019})}\BibitemShut {NoStop}%
\bibitem [{\citenamefont {Li}\ \emph {et~al.}(2019)\citenamefont {Li},
  \citenamefont {Polakovic}, \citenamefont {Wang}, \citenamefont {Xu},
  \citenamefont {Lendinez}, \citenamefont {Zhang}, \citenamefont {Ding},
  \citenamefont {Khaire}, \citenamefont {Saglam}, \citenamefont {Divan} \emph
  {et~al.}}]{li2019strong}%
  \BibitemOpen
  \bibfield  {author} {\bibinfo {author} {\bibfnamefont {Y.}~\bibnamefont
  {Li}}, \bibinfo {author} {\bibfnamefont {T.}~\bibnamefont {Polakovic}},
  \bibinfo {author} {\bibfnamefont {Y.-L.}\ \bibnamefont {Wang}}, \bibinfo
  {author} {\bibfnamefont {J.}~\bibnamefont {Xu}}, \bibinfo {author}
  {\bibfnamefont {S.}~\bibnamefont {Lendinez}}, \bibinfo {author}
  {\bibfnamefont {Z.}~\bibnamefont {Zhang}}, \bibinfo {author} {\bibfnamefont
  {J.}~\bibnamefont {Ding}}, \bibinfo {author} {\bibfnamefont {T.}~\bibnamefont
  {Khaire}}, \bibinfo {author} {\bibfnamefont {H.}~\bibnamefont {Saglam}},
  \bibinfo {author} {\bibfnamefont {R.}~\bibnamefont {Divan}}, \emph {et~al.},\
  }\href@noop {} {\bibfield  {journal} {\bibinfo  {journal} {Physical Review
  Letters}\ }\textbf {\bibinfo {volume} {123}},\ \bibinfo {pages} {107701}
  (\bibinfo {year} {2019})}\BibitemShut {NoStop}%
\bibitem [{\citenamefont {Hou}\ and\ \citenamefont
  {Liu}(2019)}]{hou2019strong}%
  \BibitemOpen
  \bibfield  {author} {\bibinfo {author} {\bibfnamefont {J.~T.}\ \bibnamefont
  {Hou}}\ and\ \bibinfo {author} {\bibfnamefont {L.}~\bibnamefont {Liu}},\
  }\href@noop {} {\bibfield  {journal} {\bibinfo  {journal} {Physical Review
  Letters}\ }\textbf {\bibinfo {volume} {123}},\ \bibinfo {pages} {107702}
  (\bibinfo {year} {2019})}\BibitemShut {NoStop}%
\bibitem [{\citenamefont {Meijer}\ \emph {et~al.}(1970)\citenamefont {Meijer},
  \citenamefont {Pimmelaar}, \citenamefont {Brouwer},\ and\ \citenamefont
  {Van~den Handel}}]{meijer1970some}%
  \BibitemOpen
  \bibfield  {author} {\bibinfo {author} {\bibfnamefont {H.}~\bibnamefont
  {Meijer}}, \bibinfo {author} {\bibfnamefont {L.}~\bibnamefont {Pimmelaar}},
  \bibinfo {author} {\bibfnamefont {S.}~\bibnamefont {Brouwer}},\ and\ \bibinfo
  {author} {\bibfnamefont {J.}~\bibnamefont {Van~den Handel}},\ }\href@noop {}
  {\bibfield  {journal} {\bibinfo  {journal} {Physica}\ }\textbf {\bibinfo
  {volume} {46}},\ \bibinfo {pages} {279} (\bibinfo {year} {1970})}\BibitemShut
  {NoStop}%
\bibitem [{\citenamefont {Pincini}\ \emph {et~al.}(2018)\citenamefont
  {Pincini}, \citenamefont {Fabrizi}, \citenamefont {Beutier}, \citenamefont
  {Nisbet}, \citenamefont {Elnaggar}, \citenamefont {Dmitrienko}, \citenamefont
  {Katsnelson}, \citenamefont {Kvashnin}, \citenamefont {Lichtenstein},
  \citenamefont {Mazurenko} \emph {et~al.}}]{pincini2018role}%
  \BibitemOpen
  \bibfield  {author} {\bibinfo {author} {\bibfnamefont {D.}~\bibnamefont
  {Pincini}}, \bibinfo {author} {\bibfnamefont {F.}~\bibnamefont {Fabrizi}},
  \bibinfo {author} {\bibfnamefont {G.}~\bibnamefont {Beutier}}, \bibinfo
  {author} {\bibfnamefont {G.}~\bibnamefont {Nisbet}}, \bibinfo {author}
  {\bibfnamefont {H.}~\bibnamefont {Elnaggar}}, \bibinfo {author}
  {\bibfnamefont {V.}~\bibnamefont {Dmitrienko}}, \bibinfo {author}
  {\bibfnamefont {M.}~\bibnamefont {Katsnelson}}, \bibinfo {author}
  {\bibfnamefont {Y.}~\bibnamefont {Kvashnin}}, \bibinfo {author}
  {\bibfnamefont {A.}~\bibnamefont {Lichtenstein}}, \bibinfo {author}
  {\bibfnamefont {V.}~\bibnamefont {Mazurenko}}, \emph {et~al.},\ }\href@noop
  {} {\bibfield  {journal} {\bibinfo  {journal} {Physical Review B}\ }\textbf
  {\bibinfo {volume} {98}},\ \bibinfo {pages} {104424} (\bibinfo {year}
  {2018})}\BibitemShut {NoStop}%
\bibitem [{\citenamefont {Turov}\ and\ \citenamefont
  {Kuleev}(1966)}]{turov1966coupled}%
  \BibitemOpen
  \bibfield  {author} {\bibinfo {author} {\bibfnamefont {E.}~\bibnamefont
  {Turov}}\ and\ \bibinfo {author} {\bibfnamefont {V.}~\bibnamefont {Kuleev}},\
  }\href@noop {} {\bibfield  {journal} {\bibinfo  {journal} {SOVIET PHYSICS
  JETP}\ }\textbf {\bibinfo {volume} {22}},\ \bibinfo {pages} {176} (\bibinfo
  {year} {1966})}\BibitemShut {NoStop}%
\bibitem [{\citenamefont {Shaltiel}(1966)}]{Shaltiel1966}%
  \BibitemOpen
  \bibfield  {author} {\bibinfo {author} {\bibfnamefont {D.}~\bibnamefont
  {Shaltiel}},\ }\href@noop {} {\bibfield  {journal} {\bibinfo  {journal}
  {Physical Review}\ }\textbf {\bibinfo {volume} {142}},\ \bibinfo {pages}
  {300} (\bibinfo {year} {1966})}\BibitemShut {NoStop}%
\bibitem [{\citenamefont {Abdurakhimov}\ \emph {et~al.}(2015)\citenamefont
  {Abdurakhimov}, \citenamefont {Bunkov},\ and\ \citenamefont
  {Konstantinov}}]{Abdurakhimov2015}%
  \BibitemOpen
  \bibfield  {author} {\bibinfo {author} {\bibfnamefont {L.~V.}\ \bibnamefont
  {Abdurakhimov}}, \bibinfo {author} {\bibfnamefont {Y.~M.}\ \bibnamefont
  {Bunkov}},\ and\ \bibinfo {author} {\bibfnamefont {D.}~\bibnamefont
  {Konstantinov}},\ }\href@noop {} {\bibfield  {journal} {\bibinfo  {journal}
  {Phys. Rev. Lett.}\ }\textbf {\bibinfo {volume} {114}},\ \bibinfo {pages}
  {226402} (\bibinfo {year} {2015})}\BibitemShut {NoStop}%
\bibitem [{\citenamefont {Abdurakhimov}\ \emph {et~al.}(2018)\citenamefont
  {Abdurakhimov}, \citenamefont {Borish}, \citenamefont {Bunkov}, \citenamefont
  {Gazizulin}, \citenamefont {Konstantinov}, \citenamefont {Kurkin},\ and\
  \citenamefont {Tankeyev}}]{Abdurakhimov2018}%
  \BibitemOpen
  \bibfield  {author} {\bibinfo {author} {\bibfnamefont {L.~V.}\ \bibnamefont
  {Abdurakhimov}}, \bibinfo {author} {\bibfnamefont {M.~A.}\ \bibnamefont
  {Borish}}, \bibinfo {author} {\bibfnamefont {Y.~M.}\ \bibnamefont {Bunkov}},
  \bibinfo {author} {\bibfnamefont {R.~R.}\ \bibnamefont {Gazizulin}}, \bibinfo
  {author} {\bibfnamefont {D.}~\bibnamefont {Konstantinov}}, \bibinfo {author}
  {\bibfnamefont {M.~I.}\ \bibnamefont {Kurkin}},\ and\ \bibinfo {author}
  {\bibfnamefont {A.~P.}\ \bibnamefont {Tankeyev}},\ }\href@noop {} {\bibfield
  {journal} {\bibinfo  {journal} {Phys. Rev. B}\ }\textbf {\bibinfo {volume}
  {97}},\ \bibinfo {pages} {0244425} (\bibinfo {year} {2018})}\BibitemShut
  {NoStop}%
\end{thebibliography}
\end{document}